\documentclass[
amsmath,
amssymb,
aps,
prl,
reprint,
superscriptaddress
]{revtex4-2}

\usepackage{bm}         
\usepackage{dcolumn}    
\usepackage{graphicx}   
\usepackage{subcaption} 
\usepackage{lipsum}     
\usepackage{xcolor}     

\usepackage{ragged2e}

\colorlet{myblue}{blue!60!black}

\usepackage[
    colorlinks=true,
    linkcolor=myblue,
    citecolor=myblue,
    urlcolor=myblue
    ]{hyperref}  

\makeatletter
\AtBeginDocument{
    \renewcommand{\NAT@open}{\textcolor{myblue}{[}}
    \renewcommand{\NAT@close}{\textcolor{myblue}{]}}
    \renewcommand{\NAT@sep}{\textcolor{myblue}{,}}
}

\def\tagform@#1{\maketag@@@{\ignorespaces#1\unskip\@@italiccorr}}
\makeatother

\begin{document}

\title{Self-Bound Droplets of Ultracold Dipolar Molecules under Tunable Double Microwave Shielding}
\author{Roger Melero}
\email{roger.melero@estudiantat.upc.edu}
\author{Jordi Boronat}
\email{jordi.boronat@upc.edu}
\author{Ferran Mazzanti}
\email{ferran.mazzanti@upc.edu}
\affiliation{Departament de Física, Universitat Politècnica de Catalunya, Campus Nord B4-B5, 08034 Barcelona, Spain}
\date{July 6, 2026}

\begin{abstract}
We use the Ground-State Path Integral Monte Carlo method to 
study a Bose-Einstein condensate of strongly interacting NaCs polar molecules under the
action of a fully anisotropic double microwave shielding potential characterized by 
a linear and an elliptical polarization field. 
In particular, we analyze the ground state of the system and its structure as a function of the 
ellipticity angle $\xi$. While for the circularly polarized case ($\xi=0$) 
a gas phase is realized, one or more self-bound droplets are observed for
small $|\xi|$'s above a threshold value near $3^\circ$. 
With increasing $\xi$, the observed droplets rapidly become tightly bound and are 
estimated to form a superfluid array. Our results compare favorably to the 
experimental observations in [Zhang et al., Nature \textbf{651}, 601 (2026)] for positive $\xi$, while 
moderate differences show up for $\xi<0$ where our simulations conform to the
expected symmetries of the intermolecular potential.

\end{abstract}

\maketitle

{\em Introduction}---Dipolar systems of atoms and molecules are of fundamental interest in quantum many-body physics because of their unique 
properties that separate them from other, more common systems naturally found in quantum ultracold gases or many-body 
condensed matter in general.
The combination of a long-range character (in three dimensions) with an anisotropic behavior~\cite{lahaye_review_2009} 
has led to a wealth of novel and intriguing phenomena, such as the formation of droplets of magnetic 
atoms~\cite{schmitt_selfbounddroplets_2016, chomaz_selfbounddroplets_2016,Baille_2020, Bottcher_2019b}, 
the realization of supersolid arrays of droplets~\cite{Guo_2019,bottcher_2019, chomaz_2019},
or the natural emergence of stripped patterns as a signature of the anisotropic character of the atomic 
interaction~\cite{Macia_2012, Lee_2024}.

While clearly successful experimental achievements with magnetic dipolar atoms have been possible, 
experimentally condensing a system of polar molecules to the quantum degeneracy regime has been elusive
due to strong recombination, together with the presence of two- and  three-body losses~\cite{dajunwang_2023, schindewolf_2025}.
These molecules, however, can present larger permanent electric dipole moments, leading to 
much stronger dipole–dipole interactions which translates into much enhanced correlations, 
unveiling new physics that are not present with magnetic atoms~\cite{Langen:2025}.
After more than two decades of effort, 
a cutting-edge setup using microwave shielding fields has been proven key to 
preventing system collapse and creating a Bose-condense ensemble of polar molecules
that lives long enough in stable form~\cite{Will:2024, Wang:2025}.

The BEC state of Ref.~\cite{Will:2024} was theoretically studied in~\cite{Sanchez:2025} 
using mean-field theory, comparing fairly well due to the small value of the gas parameter.
However, it was also shown that more correlated cases could not be properly described,
not even including Lee-Huang-Yang corrections.
The use of the path integral Monte Carlo method, which is 
an ab-initio approach to the problem, proved to be the most efficient tool to 
describe correlated ensembles of polar molecules.

\begin{figure}[ht]
\centering
\includegraphics[width=0.375\textwidth]{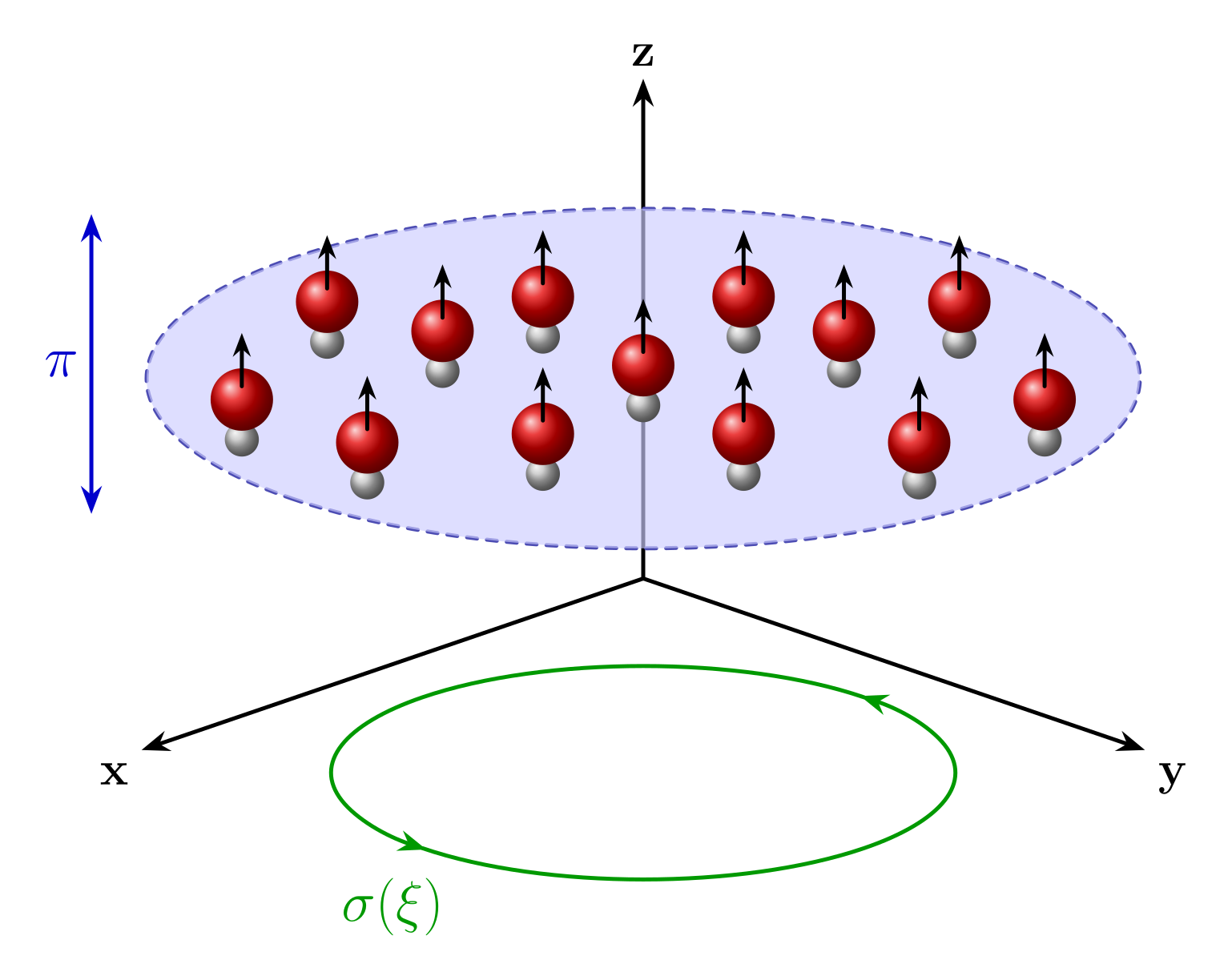}
\caption[Setup]{\justifying Schematic of the ultracold gas of NaCs molecules in an anisotropic optical dipole trap. 
The system is subjected to both a $\pi$-polarized microwave field along the $z$-axis (blue) 
and a $\sigma$-polarized field rotating in the $xy$-plane (green), parameterized by the ellipticity angle $\xi$.}
\label{fig:setup}
\end{figure}

In this Letter we study how the physics of a system of polar molecules evolve
when tuning the ellipticity of the $\sigma$-polarized microwave field 
(see Fig.\ref{fig:setup}). We use the ground-state path integral Monte 
Carlo (PIGS) method, emulating
the conditions of the NaCs experiment of Ref.~\cite{Will:2026}.
In order to to that, we use the 
interaction proposed by the authors, which
results from the balance of a dipolar 
and a van der Waals terms, both of them anisotropic. 
Our results show good agreement with the experiment for positive ellipticity angles $\xi$,
while significant differences arise for $\xi<0$.
We confirm the formation of small arrays of self-bound 
coherent clusters 
pointing to the existence of 
supersolid arrays of polar molecules.

{\em Interaction Potential}---In the novel experiment reported in Ref.~\cite{Will:2026}, the authors cooled an ensemble of sodium-cesium
(NaCs) polar molecules to the quantum degeneracy limit. To achieve this, they employed a double-shielding mechanism consisting of a combination of an elliptically ($\sigma$) and a linearly ($\pi$) polarized
microwave field, as schematically represented in Fig.~\ref{fig:setup}. 
The resulting interaction was then obtained from a perturbative expansion of a
coupled-channels scattering calculation~\cite{Karman_2025}. The intermolecular potential is well approximated
in a perturbative scheme by the sum of a purely dipolar term and a short-range part 
\begin{equation}
V({\bf r}) = V_{dd}({\bf r}) + V_{sr}({\bf r}) ,
\label{PotTot_b1}
\end{equation}
where
\begin{align}
V_{dd}({\bf r}) = \;& \frac{C_3}{r^3} \sin^2\theta\cos(2\varphi)\sin(2\xi), 
\label{PotPolMolec_a1} \\[1.5ex]
V_{sr}({\bf r}) = \;& \frac{C_6}{r^6} \Big[
1 - A_{2,0}(3\cos^2\theta-1)  \nonumber \\[0.75ex] 
& -  A_{4,0}(35 \cos^4\theta -30\cos^2\theta +3 ) \nonumber \\[0.75ex]  
& -  A_{2,2} \sin^2\theta \cos(2\varphi) \sin(2\xi) \nonumber \\[0.75ex]  
& -  A_{4,2} \sin^2\theta(7\cos^2\theta-1) \cos(2\varphi) \sin(2\xi) \nonumber \\[0.75ex] 
& - A_{4,4} \sin^4 \theta \cos(4\varphi) \sin^2(2\xi)
\Big] \ . 
\label{PotPolMolec_b1}
\end{align}
The $C_3$, $C_6$, and $A_{l,m}$ values in Eq.~\ref{PotPolMolec_b1},
corresponding to the experimental conditions, are determined from the values reported in Ref.~\cite{Will:2026}.
Using the characteristic length and energy scales $r_0=mC_3/\hbar^2, E_0=\hbar^2/mr_0^2$ 
(with $m=155.8\,{\rm u}$ for the NaCs molecule), one sets
the dimensionless $\tilde C_3 = C_3/E_0 r_0^3$ to 1 by definition. The dimensionless $\tilde C_6=C_6/E_0r_0^6$ is then
obtained from the relations $C_3=53100\sqrt{3}a_0 \hbar^2/m$ and $C_6=(3200a_0)^4\hbar^2/m$. The former indicates
that $r_0/a_0=53100\sqrt{3}$, while the latter implies that $\tilde C_6=(3200 a_0/r_0)^4=1.465\cdot 10^{-6}$. 

Notably, the $\tilde C_6$ coefficient of the short-range component of the interaction is sufficiently small 
that the whole interaction is mostly dominated by the dipolar part.
However, at very short distances of the order of $10^{-2}$ close to the minima of $V({\bf r})$,
the short-range contribution becomes comparable to the dipolar part. 
In this regime, approximating $V_sr({\bf r})$ by a 
completely isotropic term can introduce an error as large as 10\% in $V({\bf r})$ as a 
function of the direction, most affecting the behavior at $x=y=0$.
at certain directions and ellipticity angles 
compared to the fully anisotropic 
description of Eq.~\ref{PotPolMolec_b1}.

The $A_{l,m}$ constants read $A_{2,0} = 0.037, A_{2,2} = 0.153, A_{4,0} = 0.003, A_{4,2} = 0.027, A_{4,4} = 0.124$ and correspond to the ratios $(a_{6,l,m}/a_{6,0,0})^4$ obtained from the $a_{6,l,m}$ coefficients given in Ref.~\cite{Will:2026}. 
Additionally, in the experiments the system is initially confined, but the trap is afterward released to take
time-of-flight absorption images of the resulting system configurations.

\begin{figure}[ht!]
\centering
    \includegraphics[width=0.485\textwidth]{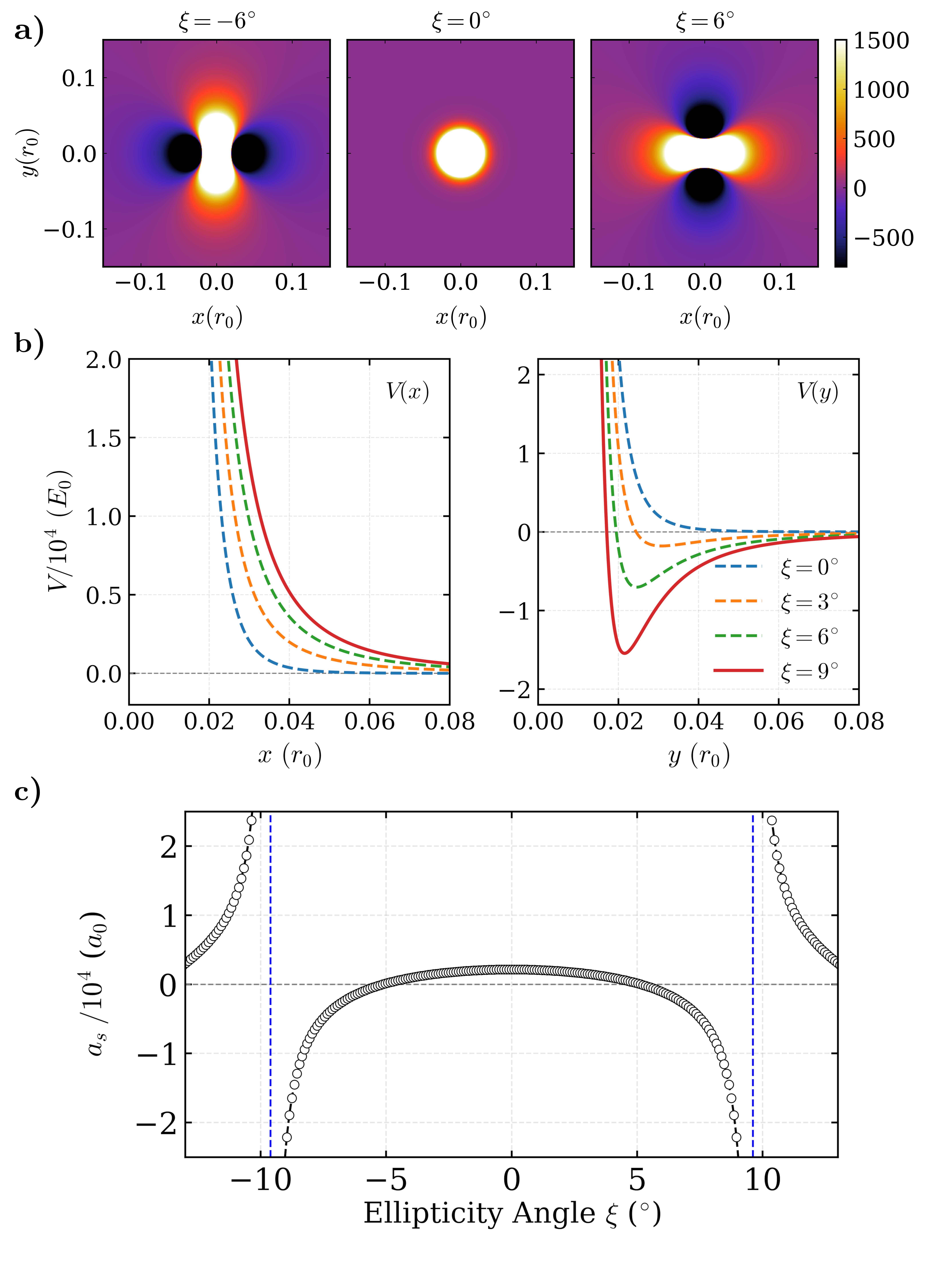}
    \caption[Potential]{ \justifying
    \textbf{a)} Intermolecular potential $V({\bf r})$ in the $xy$-plane for $\xi = -6^\circ, 0^\circ, 6^\circ$, expressed in dipolar units.
    \textbf{b)} Cuts of $V({\bf r})$ along the $x$ ($y=z=0$) and $y$ ($x=z=0$) axes for different ellipticity angles
    The transition from a purely repulsive barrier along the y-axis to the formation of a potential 
    well as $\xi$ increases is shown. 
    \textbf{c)} $s$-wave scattering length $a_s$ as a function of 
    $\xi$ for the NaCs intermolecular potential of Eqs.~\ref{PotTot_b1} to \ref{PotPolMolec_b1}
    computed using Johnson's multichannel log-derivative algorithm~\cite{Johnson:1973}. 
    The vertical dashed blue lines mark the positions of field-induced resonances.
    } 
    \label{fig:pot}
\end{figure}

Of particular relevance is the complete anisotropic character of $V({\bf r})$ induced by the polarization 
fields, as seen also in ~\cite{Deng:2023, Shi:2025, Xu:2025,deng2025two,Wang:2025}.
All these potentials display a dependence on the declination angle $\theta$, while the azimuthal angle
$\varphi$ appears only when a non-zero ellipticity ($\xi\neq0$) is considered, thereby breaking 
cylindrical symmetry. Remarkably, for $\xi=0$, the dipolar term in Eq.~\ref{PotTot_b1} vanishes, 
leaving only the short-range contribution.
These dependencies make the interaction present symmetries that affect the 
phase diagram of the system~\cite{Baillie2025DipolarMolecularGases}. 
On one hand, the interaction is invariant under the exchange $\varphi\to-\varphi$, as well as under 
the transformation $\theta\to-\theta$.
However, particularly notorious is the mapping
$\xi\to-\xi$, which corresponds to a rotation in the $xy$-plane by $\pm\frac{\pi}{2}$.
This is illustrated in Fig.~\ref{fig:pot}\textcolor{myblue}{a}, where heatmaps of the interaction $V({\bf r})$ at $z=0$ 
and $\xi=0^\circ$ and $\pm6^\circ$ are depicted.
At zero ellipticity, $V_{dd}({\bf r})$ cancels while
$V_{sr}({\bf r})$ depends only on $\theta$, which is fixed to $\pi/2$ in the plot. Therefore, the interaction depends only
on the distance to the origin. For nonzero $\xi$, however, deep attracting regions appear and rotate by $90^\circ$ with the
sign of $\xi$ as mentioned above. These regions directly affect the density profiles of the system, as discussed below.
Fig.~\ref{fig:pot}\textcolor{myblue}{b} shows cuts of $V({\bf r})$ along the $y=z=0$ and $x=z=0$ lines (left and right plots, respectively). 
As it can be seen, the interaction along the $x$-direction is always repulsive, and intensifies with increasing 
$\xi$.
The opposite happens with the cut along the $y$-axis, where the potential becomes more and more attractive the larger $\xi$ is.
At $\xi=\xi_0\sim 9.5^\circ$, the system becomes resonant as a 2-body bound state is formed~(see Fig~\ref{fig:pot}\textcolor{myblue}{c}). 
According to~\cite{Will:2026}, at this point three-body recombination losses are no
longer suppressed. 
For $\xi_0>\xi\geq 0$, the $s$-wave scattering length is positive,
and the interaction would be effectively repulsive if the system was very dilute.
However, this is not the case, as it is shown below.
In summary, the anisotropy displayed by the interaction is enhanced by an increasing ellipticity, thus favoring 
anisotropic configurations.

{\em Simulations}---In this work, we use the Path Integral Monte Carlo (PIGS) method~\cite{sarsa2000,rota2010} to simulate the ground state
of an ensemble of $N=1500$ NaCs molecules at zero temperature. In PIGS, and following Feynman's Path Integral 
formalism~\cite{feynman2018statistical}, each quantum particle 
is represented by an open  chain  of interacting coordinates (beads) that propagate from a variational model at the 
end points of the chain. At the center, unbiased configurations of the system are then obtained, provided a sufficiently 
large number of beads and an accurate short-time approximation for the imaginary-time propagator are employed. We use 
the single-parameter short-time propagator of Ref.~\cite{chin2002propagators}, which
is of order $O(\tau^6)$ for the energy and $O(\tau^4)$ for other quantities not commuting with the Hamiltonian,
with $\tau$ being the (imaginary) time step per bead. With such an accurate approximation, the number of beads required
to reach the exact ground state is greatly reduced, and the need for a high-quality trial wave function $\Phi_T$ is relaxed 
as most  of the work is done by the propagation itself. We thus use the simplest choice $\Phi_T=1$, which exactly preserves the
Bose symmetry. 

The simulations reproduce the protocol used in the experiments, where a trap is used to stabilize the system. 
We employ the same trap parameters as in Ref.~\cite{Will:2026}, corresponding to oscillator lengths
$\ell_x/r_0=0.427, \ell_y/r_0=0.318, \ell_z/r_0=0.227$. 
Once equilibrium is reached, the trap is released 
and replaced with a much wider and isotropic one, characterized by the harmonic oscillator lengths
$\ell_x/r_0 = \ell_y/r_0 = \ell_z/r_0 = 12$. Although still confined, the thermalized system 
can still be in a self-bound phase with a smaller size than the trapping lengths while 
not expelling all particles to infinity when realized as a gas.

\begin{figure}[b]
\centering
\includegraphics[width=0.4\textwidth]{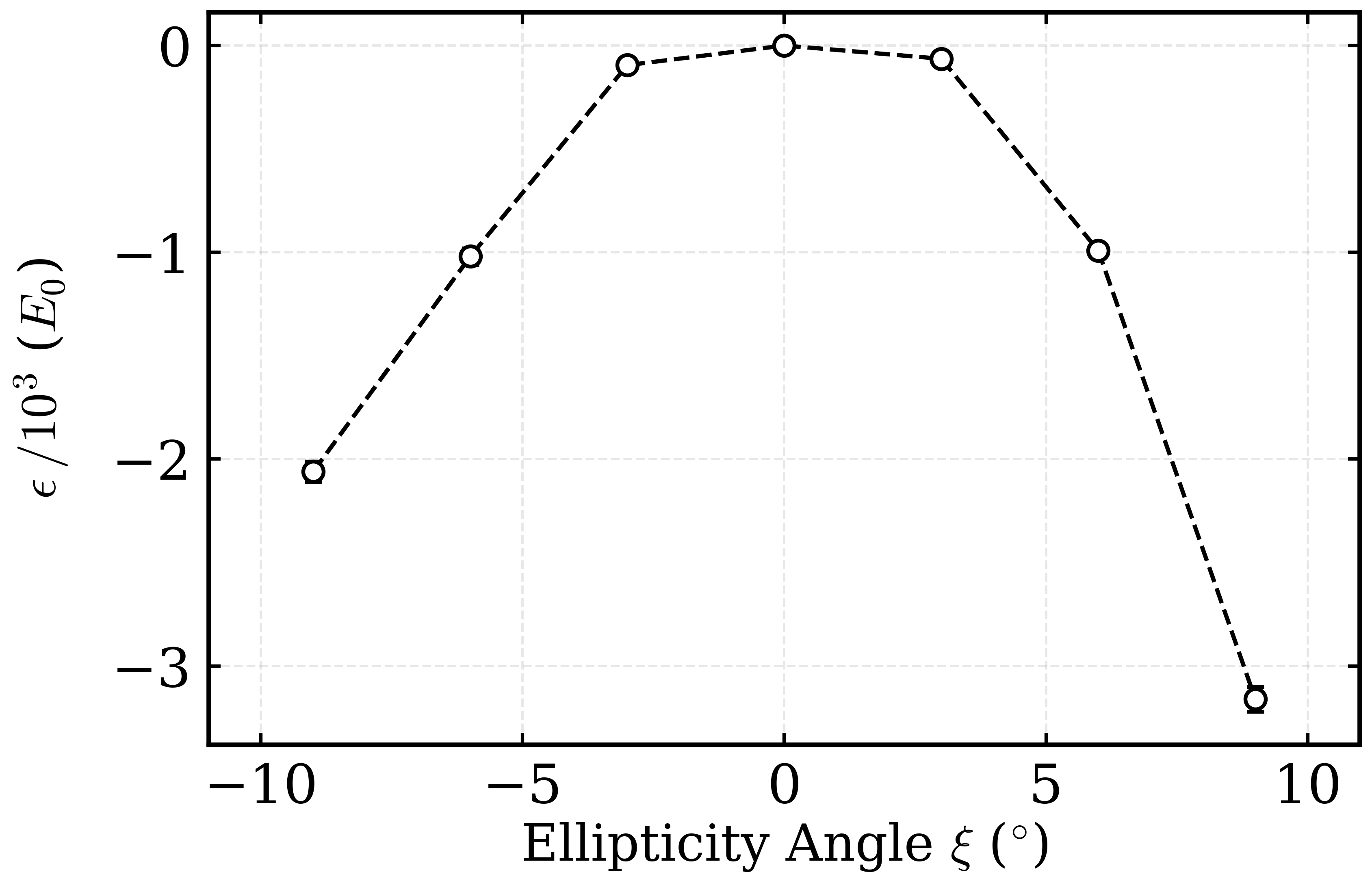}
\caption[Energy per Particle]{\justifying Energy per particle $\epsilon(\xi)$ as a function of  $\xi$ for the dressed NaCs molecular system. 
Error bars, not shown in the figure, are smaller than the size of the symbols.
At $\xi=0^\circ$ the energy is positive and achieves its maximum value, thus indicating that the system remains in gas phase.
For $|\xi|\gtrsim 3^\circ$ the energy becomes negative, forming a self-bound state.
}
\label{fig:E_N}
\end{figure}

{\em Results}---The equilibrium energy per particle $\epsilon(\xi)=E(\xi)/N$ as a function of the ellipticity parameter $\xi$ after 
releasing the trap  is shown in Fig.~\ref{fig:E_N}. As expected and according to the form of the interaction, 
$\epsilon(0)$ is non-negative, showing that the system remains in the gas phase. 
However, $\epsilon(\xi)$ rapidly decreases to very large and negative values
with increasing $\xi$, thus becoming strongly self-bound due to the large anisotropic attractive 
components arising in the interaction. As can be seen from the plot, 
$\epsilon(\xi)$ does not show a significant dependence on the sign of $\xi$ for moderate $|\xi| \lesssim 6^\circ$, 
in accordance with the corresponding symmetry displayed by $V({\bf r})$. For larger $|\xi|$, 
though, this symmetry is broken, probably due to a reminiscent dependence on the initial competition between the 
different anisotropic behaviors displayed by $V({\bf r})$ and the harmonic trap before the latter is released.

\begin{figure*}[ht]
\centering
\includegraphics[width=1.0\textwidth]{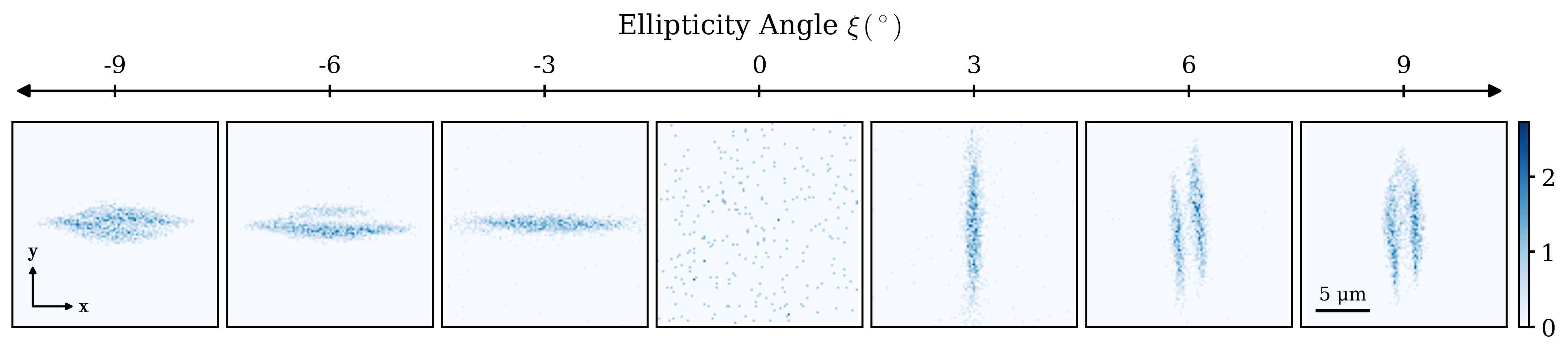}
\caption[Droplets]{\justifying 
Characteristic snapshots of an ensemble of $N=1500$ NaCs molecules, projected onto the $xy$-plane,
for different (positive and negative) ellipticity angles.
The images reveal 
the transition from a weakly dipolar BEC (at $\xi=0^\circ$) to 
a single elongated droplet and to an array of droplets as $|\xi|$ increases.
}
\label{fig:droplets}
\end{figure*}

Snapshots of the thermalized system projected onto the $xy$-plane, and for different ellipticity angles $\xi$, 
are shown in Fig.~\ref{fig:droplets}, ranging from 
$\xi=-9^\circ$ up to $\xi=+9^\circ$. The central panel of the figure shows that, for 
a perfectly circular polarization of the $\sigma$-field ($\xi=0^\circ$), the system forms a weakly dipolar 
BEC and remains in a gas phase, as
expected from the fully repulsive nature of the interaction for a vanishing ellipticity angle. However, it tends to
form a self-bound state already for small values of $\xi$, as clearly visible in the figure for $\xi=\pm 3^\circ$.
This state can be either a single droplet or an array of droplets, depending on $\xi$.
In any case, in the self-bound phase, the system organizes in prominently elongated droplets, with the major axis following the 
direction of maximal attraction of the interaction. Due to the $\xi \to -\xi$ symmetry discussed above, these arrangements
look the same but rotated by $90^\circ$ as expected, while also preserving the $\varphi \to -\varphi$ symmetry.

The formation and observed number of droplets depends on $\xi$, following the same scheme experimentally 
observed in Ref.~\cite{Will:2026}.
For very small $|\xi|$ in the $2.5^\circ$ to $3^\circ$ range, 
a single droplet is formed, while starting from a somewhat larger value $|\xi|<6^\circ$, two 
well-differentiated droplets are visible. We have not observed more than two droplets in any of the simulations 
performed once the initial tight trap is relaxed. This is in contrast with the reported absorption images obtained in 
the experiment, where after 25 ms time-of-flight measurements clearly resolve three droplets for $\xi\gtrsim 4.7^\circ$. 
This difference could be attributed to the uncertainty in the total number of molecules used in the
experiment, although this seems to be fairly unlikely.
A more fundamental and striking difference is the breakdown of the $\xi \to -\xi, \varphi \to -\varphi$ symmetries 
of the interaction, which are not observed in the experimental images, where the $+\xi$ and $-\xi$ cases show three droplets
and one single droplet, respectively. Nevertheless, in the absence of a trapping potential and with no further 
source of anisotropy, these symmetries are expected to hold unless the system is caught in a metastable state
different from the actual ground state. As a final remark, the length of the experimentally observed droplets,
measured from their major axis, is very similar to the ones obtained in the simulation for positive $\xi$.
This is not the case for the perpendicular direction, as we observe more elongated droplets than those 
reported in Ref.~\cite{Will:2026}. 
For $\xi<0$, though, our results 
show marked differences with the experiment,
as in the latter case a single droplet is observed in the whole range of ellipticity angles covered, 
thus breaking the symmetries displayed by the intermolecular potential.
Once again, the differences between simulation and experiment, and between the 
$+\xi$ and $-\xi$ configurations, could be a manifestation of the original trap used to bring the system to equilibrium, which
could at the end drive the system to a metastable state.

\begin{figure}[ht]
\centering
\includegraphics[width=0.4\textwidth]{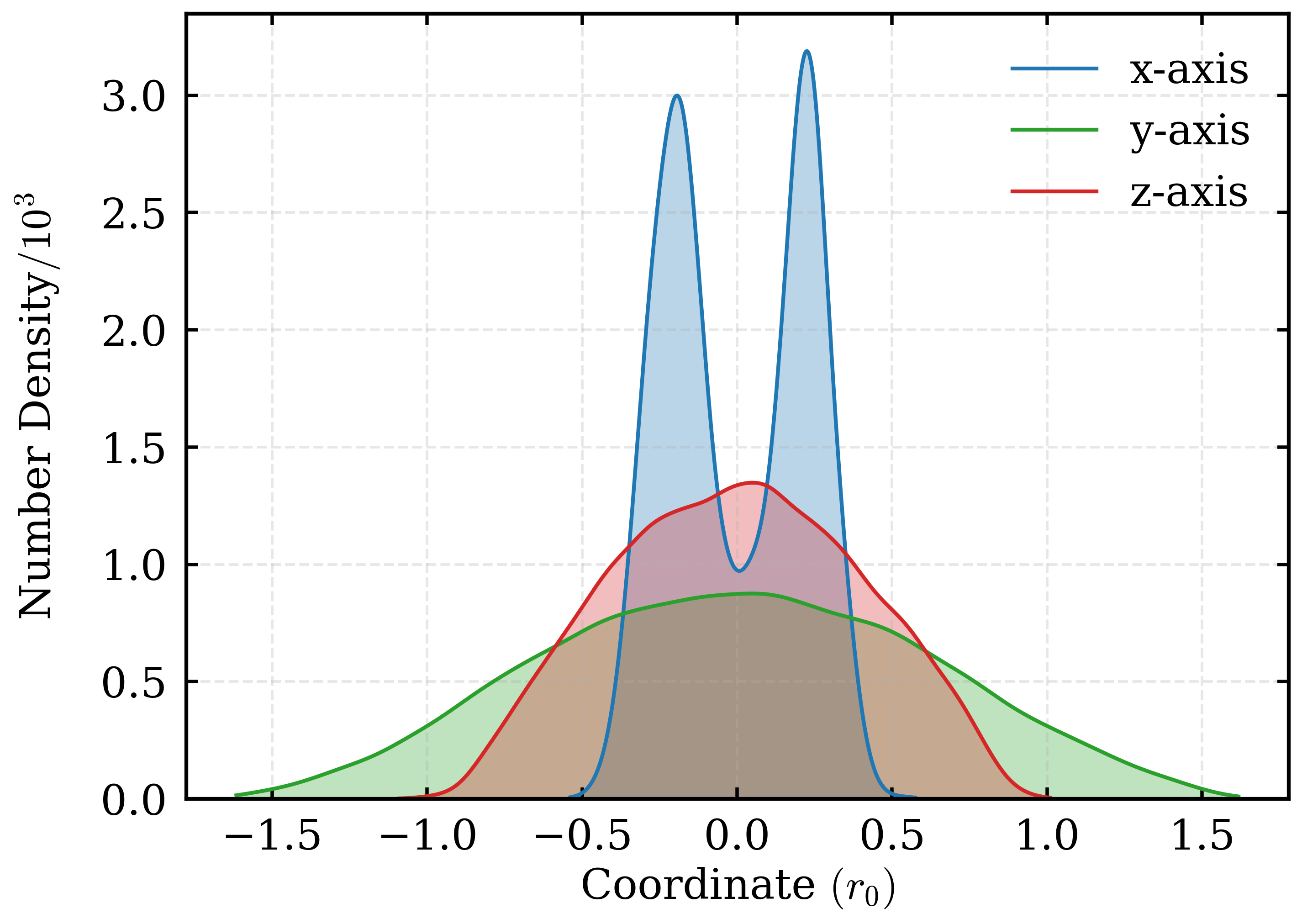}
\caption[Number Density]{\justifying One-dimensional density profiles for ellipticity angle $\xi=9^\circ$ along the $x$ (blue), $y$ (green), 
and $z$ (red) axes, expressed in dipolar units. 
}
\label{fig:num_density}
\end{figure}

We end the analysis discussing the column densities along the $x$, $y$, and $z$ directions, defined as
$n(x_i) = \int d{\bf x}_i \,n({\bf x})$, where $d{\bf x}_i$ indicates integration along all axes but the $i$-th one.
As an example, Fig.~\ref{fig:num_density} displays $n(x), n(y)$ and $n(z)$ for the $\xi=9^\circ$ case. As can be seen, there are
significant differences between the three of them, the most relevant one being that $n(x)$ presents a bimodal modulation while
the other two look almost Gaussian. The two peaks in $n(x)$ correspond to the two observed droplets of Fig.~\ref{fig:droplets}.
At the same time, the fact that the intermediate density between them is not vanishing (thus implying 
a contrast lower than 1) indicates that these droplets are not isolated and that 
quantum coherence between them can actually occur. Most relevantly, the produced array of droplets is 
stable even in the absence of an external trapping potential, in contrast to what happens with systems
of magnetic dipolar atoms, which evaporate when the trap is released. If the resulting array of droplets was
superfluid, that could be the first realization of a true supersolid of droplets.
An upper bound for the superfluid fraction $f_s$ can be obtained using Leggett's 
relation~\cite{Leggett_bounds_1543, leggett1998superfluid,perez2025superfluid}, although it is 
difficult to evaluate it accurately. For $\xi=9^\circ$, we obtain roughly $f_s \leq (0.8 \pm 0.1)$, which
is large enough to hint at the possibility of the system being superfluid. This is also in agreement with a much 
simpler superfluidity estimation based on the contrast (approximately 0.51 for the profile shown in Fig.~\ref{fig:num_density}), 
although the latter is more a measure of particle overlap 
rather than a direct measure of a purely superfluid behavior.



{\em Conclusions}---In summary, we have performed Path Integral Monte Carlo simulations of an ensemble of 
NaCs polar molecules under the same conditions 
reported in the experiment of Ref.~\cite{Will:2026}. After initial thermalization using the experimental trap, the system is released 
into a much wider confinement to reach equilibrium while preventing the complete evaporative loss of molecules.
Despite the complex anisotropic structure of both the short-range and dipolar components of the intermolecular potential, 
under the specific experimental conditions, the interaction is dominated by the dipolar part.
In this way, after the trapping stages, 
the system is expected to realize the spatial symmetries 
of the interaction in the ground state once equilibrium is reached. 
While this is not observed in the experiments, our simulations show that this is indeed the case and that 
the system becomes self-bound already for 
very small ellipticity angles (in absolute values) between $2.5^\circ$ and $3^\circ$.
For larger $|\xi|$, the energy per particle decreases steeply, driving the system into a deeply bound regime 
that leads to the formation of one or more 
droplets, exhibiting superfluid behavior as estimated using 
Leggett's upper bound for the superfluid density.
These findings pave the way for the realization of a truly self-bound supersolid state, a long-sought milestone in dipolar systems.

{\em Acknowledgments}---F.M. and J.B. acknowledge  financial support from Ministerio 
de Ciencia e Innovaci\'on MCIN/AEI/10.13039/501100011033 (Spain) under 
Grant No. PID2020-113565GB-C21.

\bibliographystyle{apsrev4-2} 
\bibliography{references}     

\end{document}